\documentclass[conference]{IEEEtran}
\IEEEoverridecommandlockouts
\usepackage{cite}
\usepackage{amsmath,amssymb,amsfonts}

\usepackage{algorithmic}
\usepackage{algorithm}
\usepackage{graphicx}
\usepackage{textcomp}
\usepackage{xcolor}

\usepackage[center]{caption} %To put caption in center in figure or table
\usepackage{subcaption}
\def\BibTeX{{\rm B\kern-.05em{\sc i\kern-.025em b}\kern-.08em
		T\kern-.1667em\lower.7ex\hbox{E}\kern-.125emX}}
\begin{document}
	
	\title{Occupant's Behavior and Emotion Based Indoor Environment's Illumination Regulation
		 \\
		%{\footnotesize \textsuperscript{*}Note: Sub-titles are not captured in Xplore and should not be used}
		%\thanks{Identify applicable funding agency here. If none, delete this.}
	}

	\author{\IEEEauthorblockN{Shreya Das}
		\IEEEauthorblockA{\textit{Department of Applied Physics} \\
			\textit{University of Calcutta}\\
			Kolkata, India \\
			shreya.das393@gmail.com}

	}

	\maketitle
	
	\begin{abstract}
		This paper presents an efficient approach for building occupancy modeling to reduce energy consumption. In this work, a novel approach to occupancy modeling based on the posture and comfort level of the occupant is developed, and subsequently, we report a new and efficient framework for detecting posture and emotion from skeleton joints and face points data respectively obtained from the Kinect sensor. The proposed approach is tested in terms of accuracy, region of convergence, and confusion matrix using several machine learning techniques. Out of all the techniques, random forest classifier gave the maximum blind test accuracy for multi-class classification of posture detection. Deep learning is used for emotion detection using several optimizers out of which Adadelta gave the maximum blind test accuracy for multi-class classification. Along with the Kinect sensor, several other sensors such as the magnetic door sensor, pyroelectric sensors, and illumination sensors are connected through a wireless network using Raspberry Pi Zero W. Thus creating an unmanned technique for illumination regulation.
	\end{abstract}
	
	%%==========================================================================================
	
	\begin{IEEEkeywords}
		Bearing-only tracking,  target motion analysis, state constrain, Lagrange multiplier.
	\end{IEEEkeywords}
	
	%%=================================================================================================================
	
	\section{Introduction}
	The ongoing scenario of total energy consumption in the world has been dominated by building energy utilization. 40\% of world’s total energy usage is used by buildings \cite{chen2015modeling}. In this age of energy impoverishment, depleting fossil fuels is making the energy needs of the 7.7 billion population across the world unattainable. Saving electricity is the most inexpensive solution to energy shortages. So constructing energy-efficient buildings has now become a major priority. The most crucial variable for this that needs constant monitoring is occupancy \cite{liao2010integrated}. Occupancy refers to the number of people present inside a space that affects the cooling or heating load, ventilation load, and illumination load to name a few, of the particular space.
	
	Several research works have been done to build efficient occupancy modeling that started with a collection of a huge amount of data using sensors. The Mitsubishi electric research labs collected data for one year using over 200 motion sensors providing over 30 million raw data to check the occupancy of two floors \cite{wren2007merl}.  Thus providing a schedule of occupancy for a year. But the occupancy modeling based on data collected over a fixed schedule often becomes outdated as the pattern of occupancy changes. The use of such a technique will also have no provision for early and late occupants. So to achieve energy efficiency real-time sensing and estimation-based control are required on all the electrical appliances that are present inside a building \cite{erickson2009energy}.
	
	Also, the use of only motion sensors gives a vague idea of the number of people present inside a room. So several other sensors are used in the literature including thermal sensors \cite{moon2010ann}, CO2 sensors \cite{arief2017cd}, cameras \cite{erickson2009energy}, pyroelectric infrared  (PIR) sensors \cite{maaspuro2018infrared}, and many more \cite{bahl2000radar, gu2009survey, liu2007survey, domdouzis2007radio}. Unfortunately, almost every sensor comes up with a drawback. The estimation of the number of people in a crowd is not possible by just using thermal sensors. The response of CO2 sensors is not in real-time and their accuracy is also reduced in properly ventilated rooms. The use of camera images needs expensive computers for image processing which are also less accurate for moving objects. The use of a camera is often debarred as it invades privacy. A PIR sensor is a binary sensor to detect motion and the use of only a PIR sensor does not provide information on the number of people present inside a room. Thus it is better to use a network of several types of sensors to obtain a better view of the number of occupants inside a room.
	
	Here, we have used real-time data obtained from various sensors enabled by wireless sensor network (WSN) \cite{vujovic2014raspberry, ferdoush2014wireless} established with the help of Raspberry Pi Zero W. Instead of using camera images we used Kinect sensor XBOX 360 that collected Cartesian coordinate data of skeleton joints and face points in a human body. This reduces the cost of image processing and also nullifies the possibility of phantom detection which is a major disadvantage of a low-cost thermal sensor. Recording just the coordinate data also does not risk the privacy of the occupants. Along with the Kinect sensor, other sensors used were door sensors, PIR sensors, and illumination sensors.
	
	Occupancy modeling based on the number of people present inside the room is commonly found in the literature \cite{yang2014systematic, munir2017real} but occupancy modeling based on the posture and comfort level of the occupant is by far new to the best of my knowledge and does not exist in literature. In this paper, we have classified posture into three classes ‘standing’, ‘sitting’, and ‘lying down'. Emotion is also classified into three classes ‘comfortable’, ‘neutral’, and ‘uncomfortable’. Illumination of the room is regulated based on these two factors. We also present a new and efficient framework for detecting posture and emotion. This includes extracting relevant features required for machine learning. 
	
	In this paper, we provide rigorous testing of posture and emotion prediction and compare the results of all the existing machine learning techniques based on both cross-validation and blind test accuracy, confusion matrix, and region of convergence (ROC) plots. Both 2-class and multi-class classification is done and results are compared. The random forest classifier gave the maximum blind test accuracy of 98.27\% for multi-class classification of posture detection. We have used deep learning for emotion detection using different optimizers out of which maximum blind test accuracy was obtained using the Adadelta optimizer which is $94.13\%$ for multi-class classification of emotion. Finally, we used the multiclass classification technique with RFC for posture detection and Adadelta optimizer for emotion detection to regulate illumination.
	The Cartesian coordinate data obtained from the Kinect sensor is not used directly to predict posture or emotion. The data for both skeletal joints and face points go through some mathematical evaluations that include essential features selection is stated in the next section of mathematical background. The proposed scheme for illumination regulation along with the different types of sensors used for data acquisition is stated in section III. The posture and emotion prediction is done using several machine learning techniques whose accuracy, confusion matrix, and region of convergence (ROC) are compared in section IV. Also, the algorithm followed to regulate the light intensity in the room is provided in the form of a flowchart in this section. Finally, the paper ends with a brief conclusion and the future scope of this work.
	\section{Mathematical Background}
	The Cartesian coordinate position of skeleton joints and face points obtained from the Kinect sensors are not used directly to determine the posture and facial expression.  Only essential features are selected from the Cartesian coordinate data positions obtained from the sensor. This also reduces data dimension and thus reduces computation complexity. The Cartesian coordinate data go through the following mathematical evaluations before they are used for machine learning.
	
		\begin{figure}[h!]
		\centering
		\includegraphics[width=40mm]{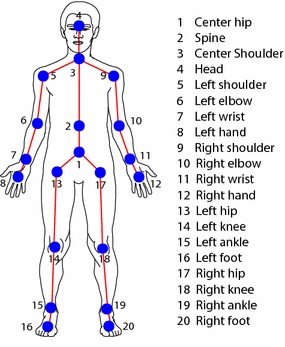}
		\caption{20 skeletal joints XBOX 360 can detect.}
		\label{fig_SkelJoints}
	\end{figure}

\begin{figure}[h!]
	\centering
	\includegraphics[width=50mm]{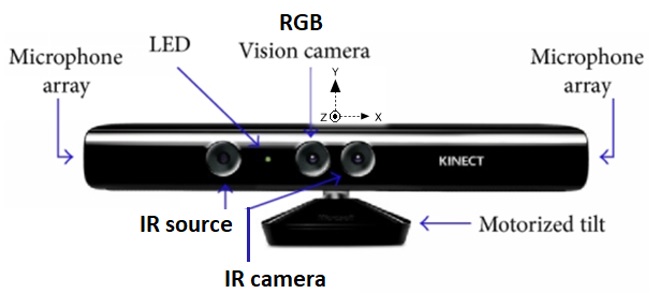}
	\caption{Kinect XBOX 360.}
	\label{fig_XBOX}
\end{figure}

\subsection{Skeleton joint data collection}
Kinect XBOX 360 is used to predict the posture of an occupant by determining the 3-dimensional Cartesian coordinate position of 20 skeleton joints of the occupant \cite{vujovic2014raspberry, ferdoush2014wireless} as shown in Figure \ref{fig_SkelJoints} with the camera center as the origin.
To classify human posture into three classes of standing, sitting, and lying down, the third-degree joints i.e. the hands and feet as shown in the Figure \ref{fig_SkelJoints} are considered to be redundant features and thus they are eliminated. To make the data independent of camera position in the room the origin is shifted from the camera center to the spine joint. Considering $(X_i, Y_i, Z_i)$ as the coordinate position of $i$-th joint with the camera as origin, $(X_s, Y_s, Z_s)$ as the position coordinates of spine joint with respect to camera center as the origin. The convention that the Kinect sensor follows for the $X$, $Y$ and $Z$ axes are shown in Figure \ref{fig_XBOX}.
Then,
\begin{equation}
	(X_{is},Y_{is},Z_{is})=(X_i,Y_i,Z_i)-(X_s,Y_s,Z_s),
\end{equation}
where $(X_{is},Y_{is},Z_{is})$ represents the coordinates of $i$-th joint with the spine joint as the origin.

The data collection for posture is done using the following logic. First, the Cartesian coordinate system is converted into a spherical coordinate system such that,
\begin{equation}
	r=\sqrt{X_{is}^2+Y_{is}^2+Z_{is}^2},
\end{equation}
\begin{equation}
	\theta=\tan ^{-1}\dfrac{\sqrt{X_{is}^2+Z_{is}^2}}{Y_{is}},
\end{equation}
and
\begin{equation}
	\phi=\tan^{-1}\dfrac{X_{is}}{Z_{is}},
\end{equation}
	where $r\in \mathbb{R}$ is the radial distance between the spine and the joints, $\theta\in \mathbb{R}$ is the angle made by the joints with $Y_{is}$ axis and $\phi\in \mathbb{R}$ is the angle made by the joints with the $Z_{is}$ axis.
	
	The value of $r$ is constant for the first degree joints like the knees, the shoulder center etc. Whereas, for second degree joints like ankles, elbows etc., the considered posture of standing, sitting and lying down is independent of $r$. Thus, the value of $r$ is neglected as it does not provide any information about posture. 
	
\begin{figure}[h!]
		\centering
		\includegraphics[width=40mm]{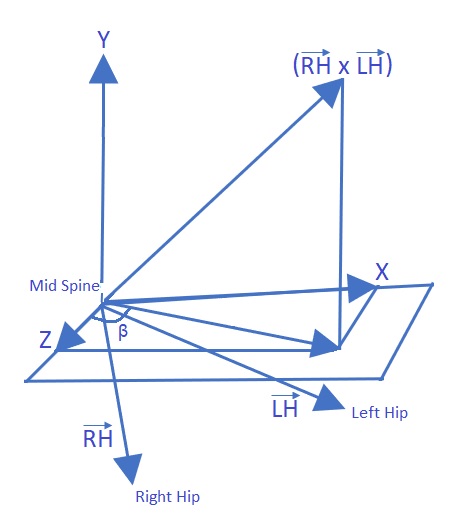}
		\caption{Vector diagram of joints with respect to the spine as origin.}
		\label{fig_vector}
	\end{figure}

	Finally, the angle of turn with respect to the $Y$-axis, $\beta\in\mathbb{R}$ as shown in Figure \ref{fig_vector} is calculated.

In Figure \ref{fig_vector}, $\vec{RH}$ represents the vector from the spine to the right-hip. $\vec{LH}$ represents the vector from spine to the left-hip. Angle of turn along $X$-axis and $Z$-axis is not needed as human body cannot physically turn or tilt along these two axes. Human body can bend, folding joints, along these axes and bending is considered by $\theta$ and $\phi$.

Thereby, we are considering a $31$-dimensional feature set vector of $15$ joints, such that, $\vec{F_{Joint}}=(\theta_1,\phi_1,\theta_2,\phi_2, ..., \theta_{15},\phi_{15},\beta )\in \mathbb{R}^{31}$ for predicting posture.

\begin{figure}[h!]
	\centering
	\includegraphics[width=10mm]{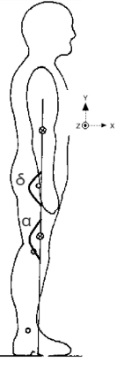}
	\includegraphics[width=10mm]{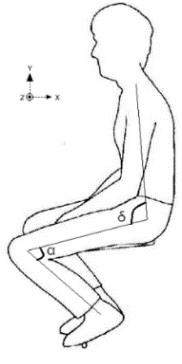}
	\includegraphics[width=30mm]{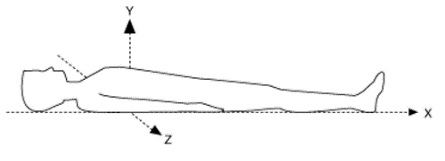}
	\centerline{(a) ~~~~~~(b)~~~~~~~~~~~~~(c)~~~~~~~~~~~}
	\caption{Posture for (a) standing, (b) sitting and (c) lying down.}
	\label{fig_Posture}
\end{figure}

Human posture of the occupant is classified into three class i.e. standing, sitting and lying down. The data labeling for posture determination is done with the logic as shown in Figure \ref{fig_Posture}.
\begin{itemize}
	\item •	From Figure \ref{fig_Posture}, if $\alpha>140^o$, $\delta>140^o$ and the torso is nearly aligned along $Y$-axis then the person is standing.
	\item If $\alpha<140^o$ and $\delta<140^o$, the person is sitting.
	\item If the torso is on the $XZ$-plane then the person is lying down.
\end{itemize}

The variables $\alpha$ and $\delta$ such that $\alpha\in\mathbb{R}$ and $\delta\in\mathbb{R}$
are calculated as,
\begin{equation}
	\alpha=\cos^{-1}\dfrac{\vec{KH}\vec{KA}}{|\vec{KH}||\vec{KA}|},
\end{equation}
	where 
	\begin{equation}
		\vec{KH}=(\vec{RH}-\vec{RK}) \text{ or }(\vec{LH}-\vec{LK}),
	\end{equation}
\begin{equation}
	\vec{KA}=(\vec{RA}-\vec{RK})\text{ or }(\vec{LA}-\vec{LK}),
\end{equation}
and
\begin{equation}
	\delta=\tan^{-1}\dfrac{\sqrt{X_{Kis}^2+Z_{Kis}^2}}{Y_{Kis}},
\end{equation}
where $(X_{Kis},Y_{Kis},Z_{Kis})$ represents the knee coordinate with respect to the spine as origin. Thus, $\delta$ is the spherical angle, $\theta$ constructed by   and   with the $Y_{is}$ axis.
The vectors used in the above equations are:
\begin{align*}
	\vec{RH}=&\text{ vector to right hip joint from spine joint,}\\
	\vec{LH}=&\text{ vector to left-hip joint from spine joint,}\\
	\vec{RK}=&\text{ vector to  right-knee joint from spine joint,}\\
	\vec{LK}=&\text{ vector to  left-knee joint from spine joint,}\\
	\vec{RA}=&\text{ vector to right-ankle joint from spine joint,}\\
	\vec{LA}=&\text{ vector to left-ankle joint from spine joint.}
\end{align*}
\subsection{Face point data collection}
\begin{figure}[h!]
	\centering
	\includegraphics[width=50mm]{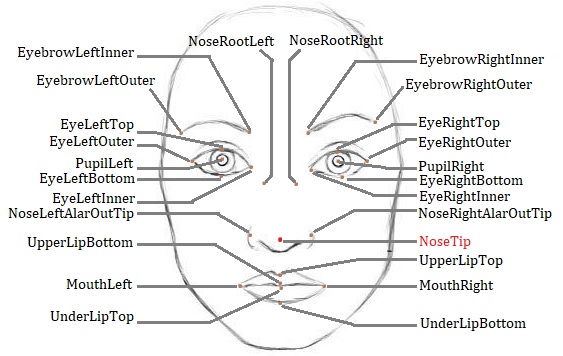}
	\centerline{(a)}
		\includegraphics[width=30mm]{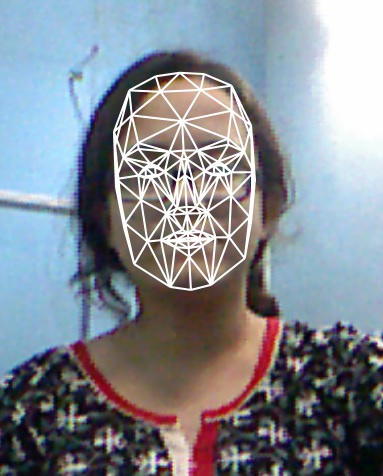}
		\includegraphics[width=50mm]{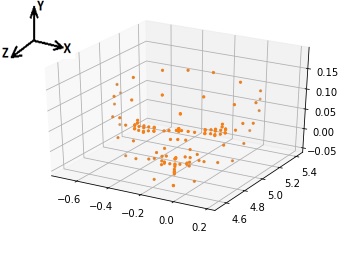}
		\centerline{(b) ~~~~~~~~~~~~~~~~~~~~~~~~~~~~~~~(c)~~~~~~~}
	\caption{A few face points as detected by Kinect XBOX 360, (b) Mesh mask of triangles whose vertices represents face points detected by Kinect XBOX 360 and (c) 3-D scatter plot of all the face points in Jupyter Notebook with scale: $X= x\times5$, $Y = y$, $Z = z\times5$ (in meter).}
	\label{fig_facePoints}
\end{figure}

Kinect XBOX 360 can determine the 3-dimensional Cartesian coordinate position of $120$ face points of human face \cite{yang2014systematic, munir2017real} with respect to the camera center as the origin, few of which is shown in Figure \ref{fig_facePoints}. Cartesian coordinate position of these $120$ face points with the nose tip as the origin is collected for our study. Assuming $(X_i, Y_i, Z_i)$ represent the coordinates of $i$-th face point with the camera as the origin and $(X_{nt}, Y_{nt}, Z_{nt})$, the position coordinates of the nose-tip face point with respect to the camera center as the origin.
Then,
\begin{equation}
	(X_{int},Y_{int},Z_{int})=(X_i,Y_i,Z_i)-(X_{nt},Y_{nt},Z_{nt}),
\end{equation}
where $(X_{int}, Y_{int}, Z_{int})$ is the coordinates of $i$-th face point with nose-tip face point as origin.
Out of these $120$ face points we have selected only $32$ face points that are highly responsible for giving an expression on human face. These $32$ face points include points in the eye-brows, mouth, cheeks and side of the nose. Cartesian coordinate position of these $32$ face points with the nose tip as the origin are separated.

\begin{figure}[h!]
	\centering
	\includegraphics[width=40mm]{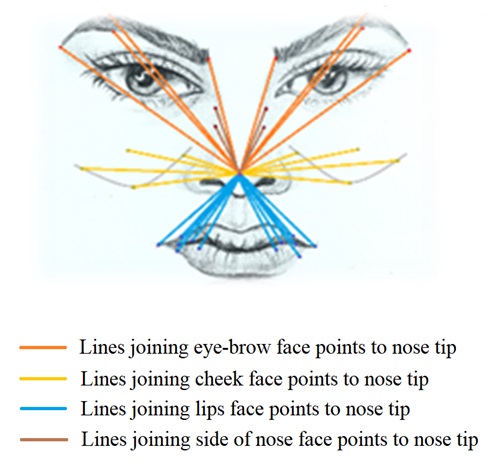}
	\includegraphics[width=40mm]{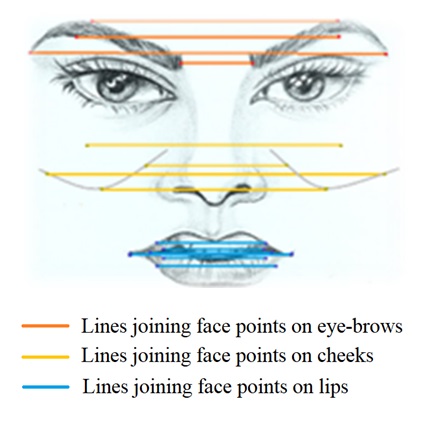}
	\centerline{(a) ~~~~~~~~~~~~~~~~~~~~~~~~~~~~(c)}
	\caption{Lines joining selected face points to the nose tip face point and (b) Lines joining selected face points to their corresponding face point on both side of face.}
	\label{fig_face}
\end{figure}

Thus, data set of $\vec{F}_{Face}  = (X_{1nt}, Y_{1nt}, Z_{1nt}, X_{2nt}, Y_{2nt}, Z_{2nt}, … X_{32nt}, Y_{32nt}, Z32nt) \in \mathbb{R}^{96}$ is obtained from the collected data set of $(X_{1nt}, Y_{1nt}, Z_{1nt}, X_{2nt}, Y_{2nt}, Z_{2nt}, … X_{120nt}, Y_{120nt}, Z_{120nt}) \in \mathbb{R}^{360}$. The data dimension is further reduced by considering the length of these face points from the nose tip and the length between the corresponding the face points on both side of the face as shown in Figure \ref{fig_face}. The Euclidean distance from the nose tip face point to the selected face points is evaluated as,
\begin{equation}
	ED_{np}=\sqrt{X_{in}^2+Y_{in}^2+Z_{in}^2},
\end{equation}
where $ED_{np}\in\mathbb{R}^{32}$ is obtained for 32 face points.

The Euclidean distance between corresponding the face points on both side of the face is evaluated as,
\begin{equation}
	ED_h=\sqrt{(X_{Rin}-X_{Lin})^2+(Y_{Rin}-Y_{Lin})^2+(Z_{Rin}-Z_{Lin})^2},
\end{equation}
where $(X_{Rin}, Y_{Rin}, Z_{Rin})$ is the Cartesian coordinate of the i-th face point on the right side of the face with the nose tip as the origin and $(X_{Lin}, Y_{Lin}, Z_{Lin})$ is the Cartesian coordinate of the i-th face point on the left side of the face with the nose tip as the origin.
$ED_h$ is obtained for 28 face points. The face points on the side of the nose is not considered here as those points have no horizontal movement while delivering an expression. So, $ED_h \in \mathbb{R}^{14}$. 
All these $ED_{np}$ and $ED_h$ makes $(32+14)=46$ features in total that are used to perform machine learning techniques to predict emotion. Thereby, the final feature space, $\phi_{fn} \in \mathbb{R}^{46}$ is obtained for emotion detection. Data labeling is done manually for emotion determination.

\section{Materials and Methods}
\subsection{Proposed Scheme}
\begin{figure}[h!]
	\centering
	\includegraphics[width=40mm]{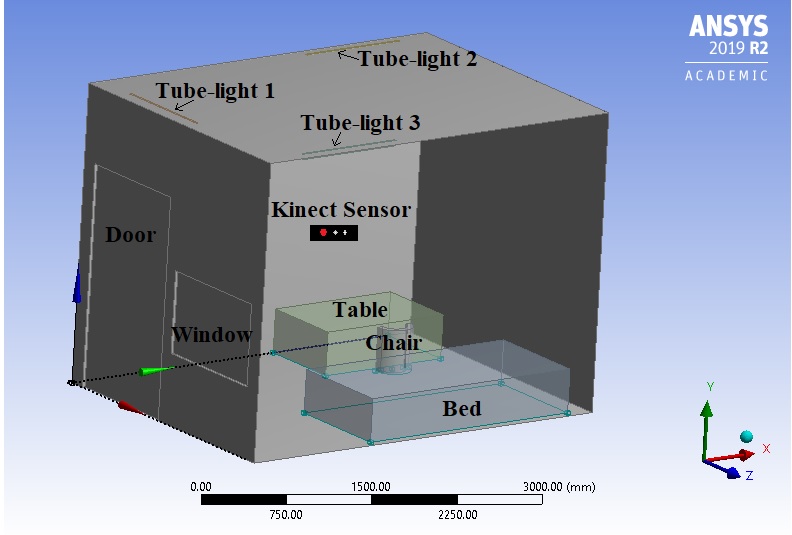}
	\caption{Prototype test bed room.}
	\label{fig_room}
\end{figure}

\begin{figure}[h!]
	\centering
	\includegraphics[width=70mm]{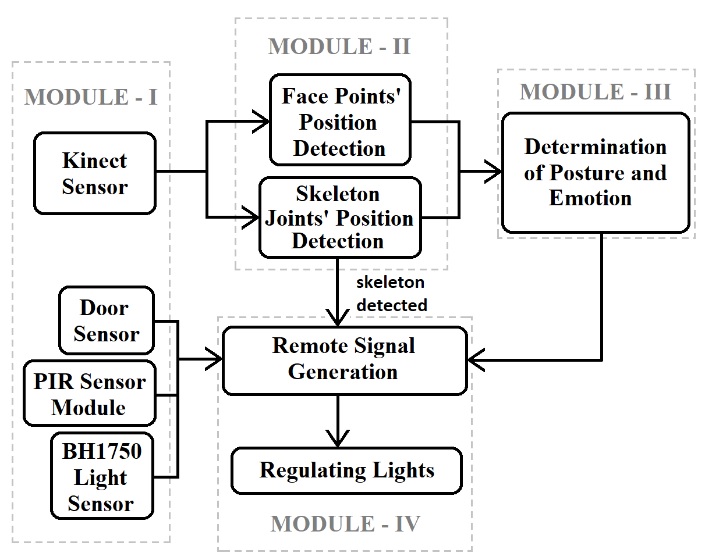}
	\caption{System schematic diagram.}
	\label{fig_scheme}
\end{figure}

The whole setup is employed in a prototype test bed room as shown in Figure \ref{fig_room}. The room has a door having a door sensor and two PIR sensors, a window, a bed, a table above which the Kinect sensor is placed and three tube lights for illumination regulation. 
The proposed scheme that is used to regulate light intensity in the test bed room as shown in Figure \ref{fig_room} depending on the posture and comfort level of the occupant inside the room is divided into four modules as shown in Figure \ref{fig_scheme}.

Each module in Figure \ref{fig_scheme} performs a different task as explained below,
\begin{itemize}
	\item Module I consists of sensors used in the system.
	\item Module II represents the determination of skeleton joints and face points from Kinect XBOX 360.
	\item Module III refers to the posture and emotion classification of the occupant.
	\item Module IV regulates tube lights’ lumen as per the comfort level of the occupant.
\end{itemize}

\subsection{Data acquisition}
Data acquisition is performed using different sensors that include the magnetic door sensor, PIR sensor, BH1750 light intensity sensor, and Kinect XBOX 360.

\subsubsection{Magnetic door sensor}
A binary reed switch magnetic door sensor is used to check the door status i.e. the door is open or closed. This status is sent over wirelessly using Raspberry Pi Zero W, to the computer where the final coding for illumination regulation is done \cite{gordon2002wireless, canceill1969magnetic}. 
\subsubsection{PIR sensor}
The binary PIR sensor detects motion as a human passes by the sensor. Two PIR sensors are fitted near the door. One of them is fitted inside and the other is fitted outside the room. When someone enters the room the PIR sensor fitted outside the room detects motion first and then the PIR sensor fitted inside the room detects motion. The reverse happens when someone leaves the room. Using this technique we can count the number of people entering and leaving the room thus we can also count the occupancy inside the room. This data is sent over to a computer wirelessly via socket communication using Raspberry Pi Zero W \cite{kaundanya2017smart, patel2016smart}.
\subsubsection{BH1750}
BH1750 is a digital ambient light sensor IC that uses I2C bus interface \cite{gao2017light, wang2011design}. Each tube light is fitted with a BH1750 sensor. The status of light intensity inside the room is obtained in real-time from BH1750 using Raspberry Pi Zero W which is again communicated to the computer.
\subsubsection{Kinect sensor (XBOX 360)}
\begin{figure}[h!]
	\centering
	\includegraphics[width=10mm]{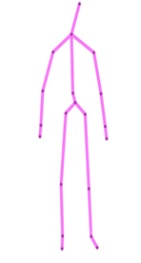}
	\includegraphics[width=10mm]{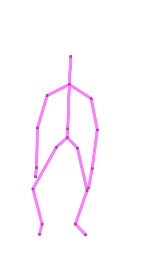}
	\includegraphics[width=30mm]{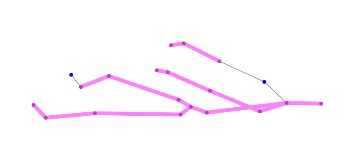}
	\centerline{(a) ~~~~~~(b)~~~~~~~~~~~~~(c)~~~~~~~~~~~}
	\caption{3 posture of (a) Standing, (b) Sitting and (c) Lying Down.}
	\label{fig_PostureBone}
\end{figure}

\begin{figure}[h!]
	\centering
	\includegraphics[width=40mm]{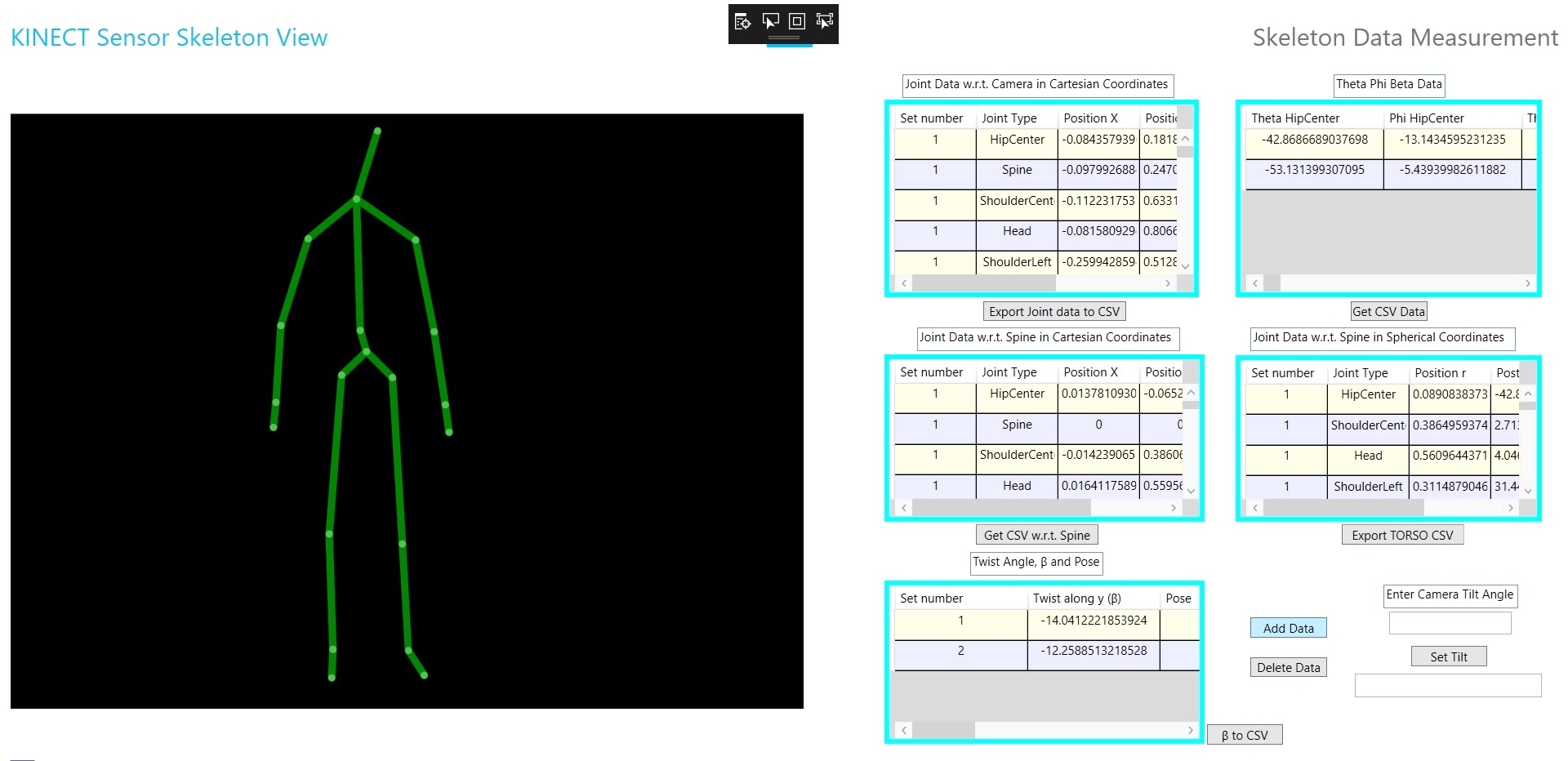}
	\includegraphics[width=39.25mm]{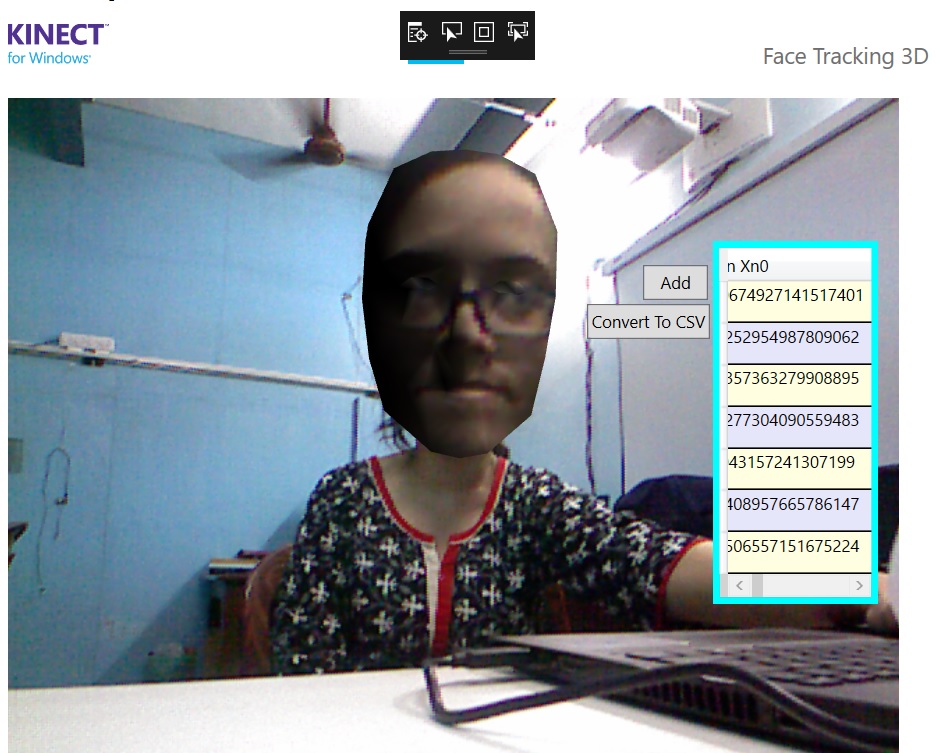}
	\centerline{~~~(a) ~~~~~~~~~~~~~~~~~~~~~~~~~~(b)}
	\caption{Skeleton joint data acquisition, (b) Face points data acquisition.}
	\label{fig_XBOXImm}
\end{figure}

The Kinect sensor detects human presence and gathers coordinate data of skeleton joints and face points. The coordinate data after going through some evaluations as discussed in the mathematical background section is collected to predict the posture i.e. `standing', `sitting' or `lying down' (as shown in Figures \ref{fig_PostureBone} and \ref{fig_XBOXImm}), and comfort level i.e. `comfortable', `neutral' or `uncomfortable'. Data set is obtained in the computer from Kinect sensor using visual studio (WPF application C\#) as shown in Figure \ref{fig_XBOXImm}.

\subsection{Illumination regulation}
\begin{figure}[h!]
	\centering
	\includegraphics[width=50mm]{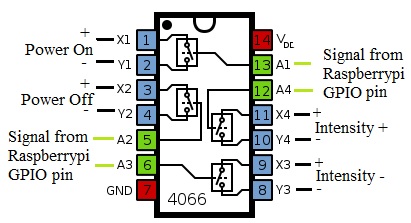}
	\caption{Circuit attached to remote controls of tube lights.}
	\label{fig_remote}
\end{figure}

Three tube lights are installed inside the test bed room whose intensity is to be regulated. Each of the tube lights has remote control with it which is used to regulate the light intensity in the room. A small circuit is built that is connected to the remote buttons as shown in Figure \ref{fig_remote}.
The terminals of `power on', `power off', `intensity +' and `intensity –' buttons of tube lights’ remote controls are soldiered to the pins of switching IC 4066 as shown in Figure \ref{fig_remote}. Control signal is sent to the pins of 4066 from Raspberry Pi to short the desired button terminals on the remote control of the tube light whose illumination is to be regulated.

\section{Results}
\subsection{Posture Determination}
\begin{figure}[h!]
	\centering
	\includegraphics[width=50mm]{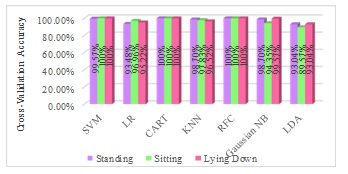}
	\caption{Cross-validation accuracy plot for 2 class classification of posture.}
	\label{fig_2ClassPosCross}
\end{figure}

\begin{figure}[h!]
	\centering
	\includegraphics[width=50mm]{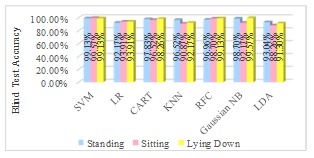}
	\caption{Blind test accuracy plot for 2 class classification of posture.}
	\label{fig_2ClassPosBlind}
\end{figure}

\begin{figure}[h!]
	\centering
	\includegraphics[width=40mm]{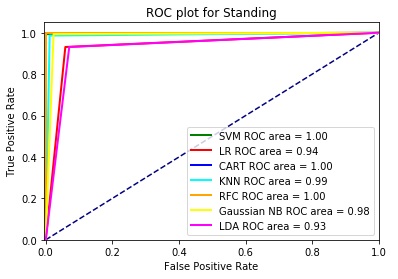}
	\includegraphics[width=40mm]{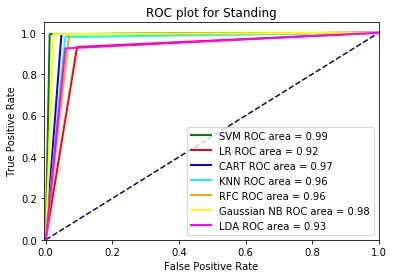}
	\centerline{~~~(a) ~~~~~~~~~~~~~~~~~~~~~~~~~~(b)}
		\includegraphics[width=40mm]{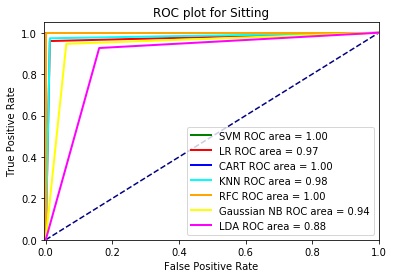}
	\includegraphics[width=40mm]{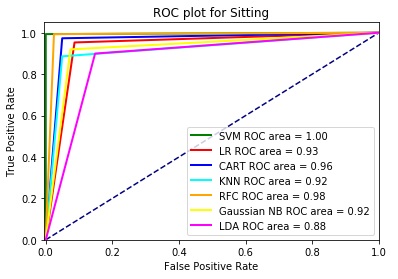}
	\centerline{~~~(c) ~~~~~~~~~~~~~~~~~~~~~~~~~~(d)}
		\includegraphics[width=40mm]{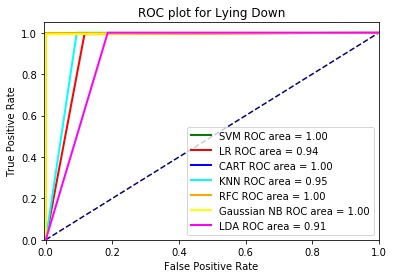}
	\includegraphics[width=40mm]{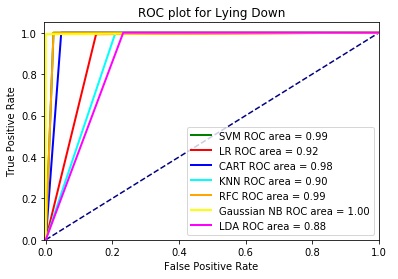}
	\centerline{~~~(e) ~~~~~~~~~~~~~~~~~~~~~~~~~~(f)}
	\caption{ROC of 2 class classification; for standing vs. rest (a) cross-validation (b) blind test; for sitting vs. rest (c) cross-validation (d) blind test and for lying down vs. rest (e) cross-validation (f) blind test to determine posture.}
	\label{fig_14}
\end{figure}

Skeleton joint data was obtained from as many as 27 people using visual studio 2019 (WPF application C\#). Several machine learning techniques are used to predict posture. The machine learning techniques used are Support Vector Machine (SVM), Logistic Regression (LR), Classification and Regression Trees (CART), K-Nearest Neighbor (KNN), Random Forest Classifier (RFC), Gaussian Naive Bayes (Gaussian NB) and Linear Discriminant Analysis (LDA). Table \ref{Tab_I}, \ref{Tab_II} and \ref{Tab_III} shows the confusion matrix for 2 class classification of standing versus rest, sitting versus rest and lying down versus rest respectively. Figure \ref{fig_2ClassPosCross} and \ref{fig_2ClassPosBlind} shows the cross-validation an blind test accuracy, respectively obtained for 2 class classification of posture. The corresponding ROC plots are provided in Figure \ref{fig_14}. 
\begin{table}[h!]
	\renewcommand{\arraystretch}{1.1}
	\setlength{\tabcolsep}{5pt}
	\centering 
	\caption{Cross validation and blind test confusion matrix for 2 class classification of posture, standing versus rest using different classifiers}
	\begin{tabular}{|c|c|c|}
		\hline
		\textbf{Classifier}&\textbf{Cross-Validation}&\textbf{Blind Test}\\\hline
		SVM   &	$\begin{bmatrix}69&0\\1&160
		\end{bmatrix}$	&	$\begin{bmatrix}
		69&0\\4&157
	\end{bmatrix}$		\\\hline

		LR   &	$\begin{bmatrix}65&4\\12&149
		\end{bmatrix}$	&	$\begin{bmatrix}
			67&2\\16&160
		\end{bmatrix}$		\\\hline
	
	CART   &	$\begin{bmatrix}69&0\\0&161
	\end{bmatrix}$	&	$\begin{bmatrix}
		67&2\\1&160
	\end{bmatrix}$		\\\hline
		
	KNN  &	$\begin{bmatrix}68&1\\8&153
		\end{bmatrix}$	&	$\begin{bmatrix}
			65&4\\16&145
		\end{bmatrix}$		\\\hline
	
	RFC   &	$\begin{bmatrix}69&0\\0&161
	\end{bmatrix}$	&	$\begin{bmatrix}
		69&0\\3&158
	\end{bmatrix}$		\\\hline

Gaussian NB  &	$\begin{bmatrix}68&1\\1&160
\end{bmatrix}$	&	$\begin{bmatrix}
	68&1\\1&160
\end{bmatrix}$		\\\hline

LDA   &	$\begin{bmatrix}68&1\\16&145
\end{bmatrix}$	&	$\begin{bmatrix}
	68&1\\18&143
\end{bmatrix}$		\\\hline
	\end{tabular}
	\label{Tab_I}
\end{table}

\begin{table}[h!]
	\renewcommand{\arraystretch}{1.1}
	\setlength{\tabcolsep}{5pt}
	\centering 
	\caption{Cross-validation and blind test confusion matrix for class 2 classification of posture, sitting versus rest using different classifiers}
	\begin{tabular}{|c|c|c|}
		\hline
		\textbf{Classifier}&\textbf{Cross-Validation}&\textbf{Blind Test}\\\hline
		SVM   &	$\begin{bmatrix}86&1\\1&142
		\end{bmatrix}$	&	$\begin{bmatrix}
			84&3\\1&142
		\end{bmatrix}$		\\\hline
		
		LR   &	$\begin{bmatrix}82&5\\10&133
		\end{bmatrix}$	&	$\begin{bmatrix}
			80&7\\13&130
		\end{bmatrix}$		\\\hline
		
		CART   &	$\begin{bmatrix}87&0\\0&143
		\end{bmatrix}$	&	$\begin{bmatrix}
			85&2\\4&139
		\end{bmatrix}$		\\\hline
		
		KNN  &	$\begin{bmatrix}84&3\\11&132
		\end{bmatrix}$	&	$\begin{bmatrix}
			82&35\\17&136
		\end{bmatrix}$		\\\hline
		
		RFC   &	$\begin{bmatrix}86&1\\0&143
		\end{bmatrix}$	&	$\begin{bmatrix}
			85&2\\1&142
		\end{bmatrix}$		\\\hline
		
		Gaussian NB  &	$\begin{bmatrix}85&2\\9&134
		\end{bmatrix}$	&	$\begin{bmatrix}
			85&2\\13&130
		\end{bmatrix}$		\\\hline
		
		LDA   &	$\begin{bmatrix}81&6\\14&129
		\end{bmatrix}$	&	$\begin{bmatrix}
			81&6\\16&127
		\end{bmatrix}$		\\\hline
	\end{tabular}
	\label{Tab_II}
\end{table}

\begin{table}[h!]
	\renewcommand{\arraystretch}{1.1}
	\setlength{\tabcolsep}{5pt}
	\centering 
	\caption{Cross-validation and blind test confusion matrix for class 2 classification of posture, lying down versus rest using different classifiers}
	\begin{tabular}{|c|c|c|}
		\hline
		\textbf{Classifier}&\textbf{Cross-Validation}&\textbf{Blind Test}\\\hline
		SVM   &	$\begin{bmatrix}68&0\\0&162
		\end{bmatrix}$	&	$\begin{bmatrix}
			67&1\\0&162
		\end{bmatrix}$		\\\hline
		
		LR   &	$\begin{bmatrix}58&10\\1&161
		\end{bmatrix}$	&	$\begin{bmatrix}
			58&10\\1&161
		\end{bmatrix}$		\\\hline
		
		CART   &	$\begin{bmatrix}68&0\\0&162
		\end{bmatrix}$	&	$\begin{bmatrix}
			67&1\\2&160
		\end{bmatrix}$		\\\hline
		
		KNN  &	$\begin{bmatrix}59&9\\0&162
		\end{bmatrix}$	&	$\begin{bmatrix}
			57&11\\0&162
		\end{bmatrix}$		\\\hline
		
		RFC   &	$\begin{bmatrix}68&0\\1&161
		\end{bmatrix}$	&	$\begin{bmatrix}
			68&0\\0&162
		\end{bmatrix}$		\\\hline
		
		Gaussian NB  &	$\begin{bmatrix}68&0\\1&161
		\end{bmatrix}$	&	$\begin{bmatrix}
			68&0\\0&162
		\end{bmatrix}$		\\\hline
		
		LDA   &	$\begin{bmatrix}55&13\\0&162
		\end{bmatrix}$	&	$\begin{bmatrix}
			55&13\\0&162
		\end{bmatrix}$		\\\hline
	\end{tabular}
	\label{Tab_III}
\end{table}

\begin{table}
		\renewcommand{\arraystretch}{1.1}
	\setlength{\tabcolsep}{2pt}
	\centering 
	\caption{Confusion matrix for multi-class classification to determine posture.}
	\begin{tabular}{|c|c|c|c|c|}\hline
		\textbf{Classifier}&\multicolumn{2}{c|}{\textbf{Cross-Validation}}&\multicolumn{2}{c|}{\textbf{Blind Test}}\\\cline{2-5}
		&\textbf{Accuracy}&\textbf{Confusion}&\textbf{Accuracy}&\textbf{Confusion}\\
		&\textbf{\%}&\textbf{Matrix}&\textbf{\%}&\textbf{Matrix}\\\hline
		RFC&100&$\begin{bmatrix}
			76&0&0\\1&82&0\\0&0&72
		\end{bmatrix}$&98.27&$\begin{bmatrix}
		73&3&0\\2&77&4\\0&0&72
	\end{bmatrix}$\\\hline

KNN&92.24&$\begin{bmatrix}
	72&4&0\\6&77&0\\4&5&63
\end{bmatrix}$&84.85&$\begin{bmatrix}
70&6&0\\10&73&0\\10&7&55
\end{bmatrix}$\\\hline

CART&99.57&$\begin{bmatrix}
	75&0&1\\0&82&1\\1&0&71
\end{bmatrix}$&97.40&$\begin{bmatrix}
75&1&0\\5&74&4\\1&1&70
\end{bmatrix}$\\\hline

LDA&94.81&$\begin{bmatrix}
	75&1&0\\4&79&0\\6&8&58
\end{bmatrix}$&88.74&$\begin{bmatrix}
75&1&0\\4&79&0\\7&10&55
\end{bmatrix}$\\\hline

LR&94.81&$\begin{bmatrix}
	76&0&0\\2&80&1\\3&4&65
\end{bmatrix}$&91.77&$\begin{bmatrix}
76&0&0\\3&77&3\\5&5&62
\end{bmatrix}$\\\hline

SVM&97.40&$\begin{bmatrix}
	76&0&0\\1&82&0\\3&4&65
\end{bmatrix}$&93.94&$\begin{bmatrix}
76&0&0\\1&82&0\\4&5&63
\end{bmatrix}$\\\hline

Gaussian NB&99.13&$\begin{bmatrix}
	74&1&1\\0&79&4\\0&0&72
\end{bmatrix}$&97.84&$\begin{bmatrix}
74&1&1\\0&77&6\\0&0&72
\end{bmatrix}$\\\hline
	\end{tabular}
	\label{Tab_IIII}
\end{table}

\begin{figure}[h!]
	\centering
	\includegraphics[width=50mm]{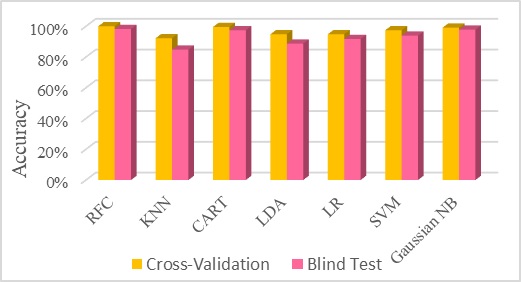}
	\caption{Multi-class classification accuracy for posture determination.}
	\label{fig_15}
\end{figure}

In case of the cross-validation and blind test accuracy plot for 2 class classification using several machine learning techniques. After performing the LDA which is a data dimension reduction technique, the SVM is performed to check accuracy. The RFC and the CART shows $100\%$ 
cross-validation accuracy and the SVM shows maximum blind test accuracy for 2 class classification. The SVM shows maximum area under the ROC (Receiver Operator Characteristics) curves as shown in Figure \ref{fig_14}.
The random forest classifier shows the maximum accuracy in both blind test and cross-validation test as shown in Figure \ref{fig_15} and Table \ref{Tab_IIII} for multi-class classification. Confusion matrices for the RFC shown in Table \ref{Tab_IIII} have  high diagonal element values thus showing accurate prediction of posture.

\begin{figure}[h!]
	\centering
	\includegraphics[width=50mm]{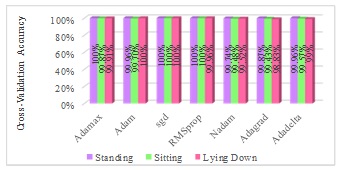}
	\caption{Cross-validation accuracy plot for 2 class classification of posture.}
	\label{fig_2ClassPosCrossANN}
\end{figure}

\begin{figure}[h!]
	\centering
	\includegraphics[width=50mm]{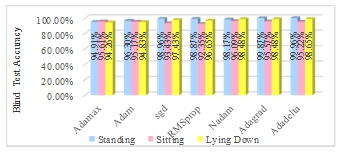}
	\caption{Blind test accuracy plot for 2 class classification of posture.}
	\label{fig_2ClassPosBlindANN}
\end{figure}

\begin{figure}[h!]
	\centering
	\includegraphics[width=40mm]{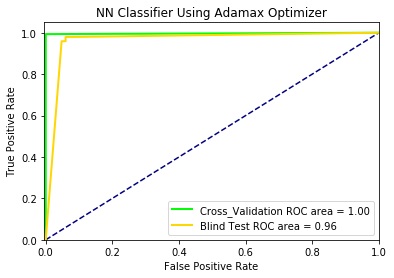}
	\includegraphics[width=40mm]{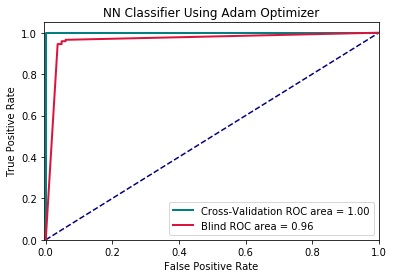}
	\centerline{~~~(a) ~~~~~~~~~~~~~~~~~~~~~~~~~~(b)}
	\includegraphics[width=40mm]{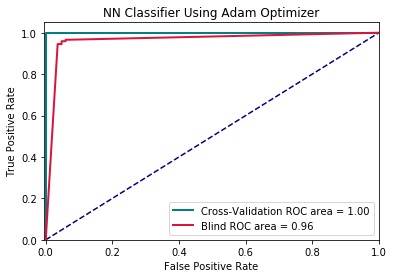}
	\includegraphics[width=40mm]{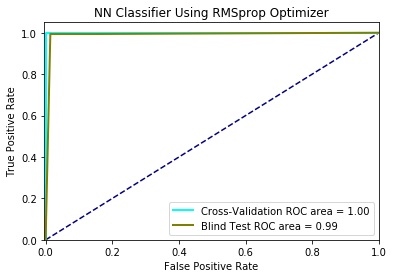}
	\centerline{~~~(c) ~~~~~~~~~~~~~~~~~~~~~~~~~~(d)}
	\includegraphics[width=40mm]{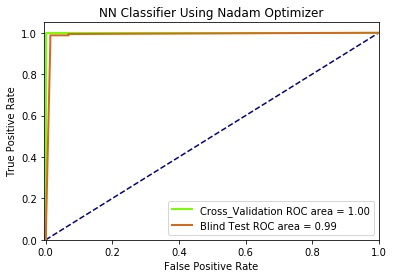}
	\includegraphics[width=40mm]{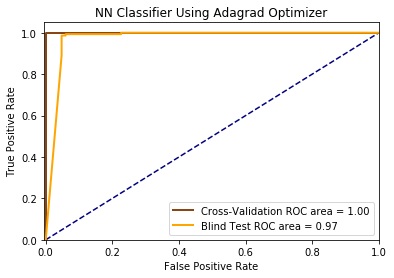}
	\centerline{~~~(e) ~~~~~~~~~~~~~~~~~~~~~~~~~~(f)}
	\includegraphics[width=40mm]{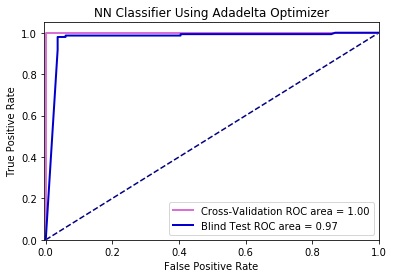}
	\centerline{(g)}
	\caption{Fig. ROC plot for 2 class classification of posture; standing versus rest using (a) Adamax, (b) Adam, (c) sgd, (d) RMSprop, (e) Nadam, (f) Adagrad and (g) Adadelta optimizers.}
	\label{fig_18}
\end{figure}

\begin{figure}[h!]
	\centering
	\includegraphics[width=40mm]{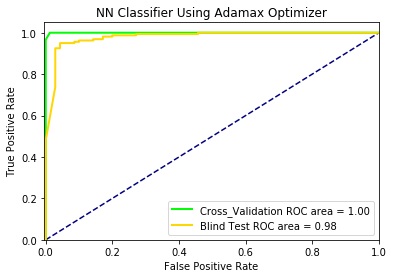}
	\includegraphics[width=40mm]{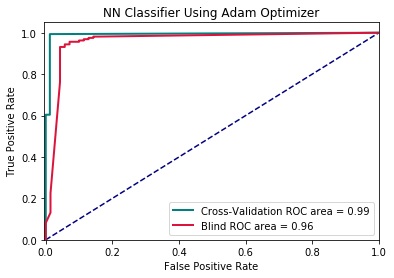}
	\centerline{~~~(a) ~~~~~~~~~~~~~~~~~~~~~~~~~~(b)}
	\includegraphics[width=40mm]{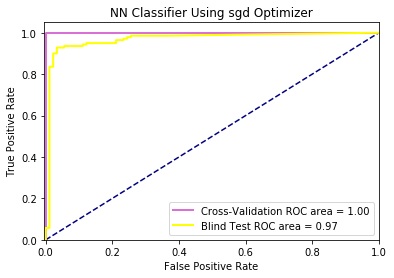}
	\includegraphics[width=40mm]{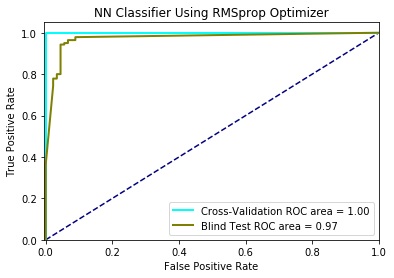}
	\centerline{~~~(c) ~~~~~~~~~~~~~~~~~~~~~~~~~~(d)}
	\includegraphics[width=40mm]{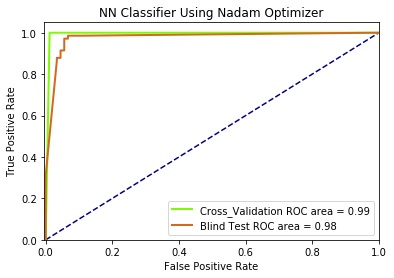}
	\includegraphics[width=40mm]{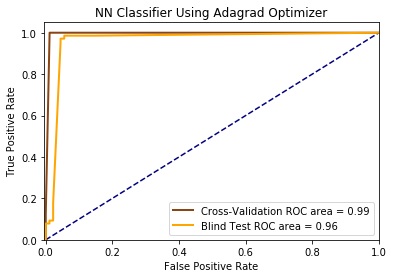}
	\centerline{~~~(e) ~~~~~~~~~~~~~~~~~~~~~~~~~~(f)}
	\includegraphics[width=40mm]{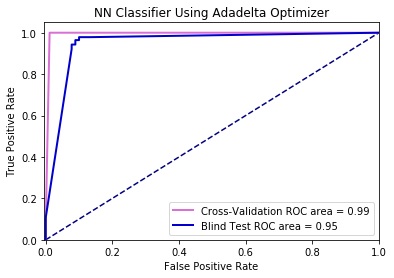}
	\centerline{(g)}
	\caption{ROC plot for 2 class classification of posture; sitting versus rest using (a) Adamax, (b) Adam, (c) sgd, (d) RMSprop, (e) Nadam, (f) Adagrad and (g) Adadelta optimizers.}
	\label{fig_19}
\end{figure}

\begin{figure}[h!]
	\centering
	\includegraphics[width=40mm]{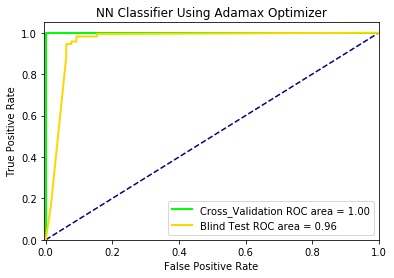}
	\includegraphics[width=40mm]{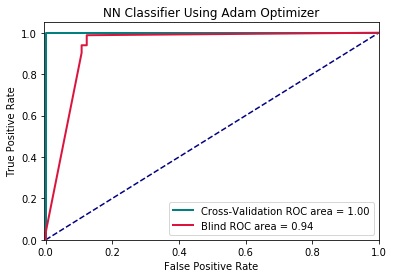}
	\centerline{~~~(a) ~~~~~~~~~~~~~~~~~~~~~~~~~~(b)}
	\includegraphics[width=40mm]{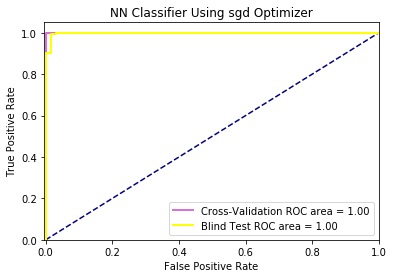}
	\includegraphics[width=40mm]{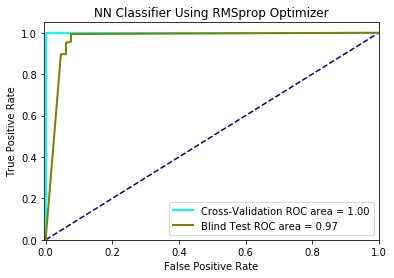}
	\centerline{~~~(c) ~~~~~~~~~~~~~~~~~~~~~~~~~~(d)}
	\includegraphics[width=40mm]{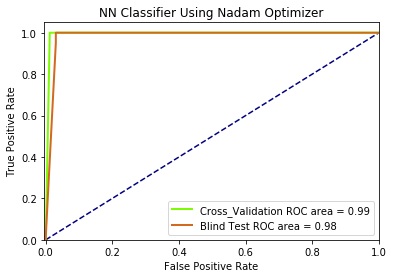}
	\includegraphics[width=40mm]{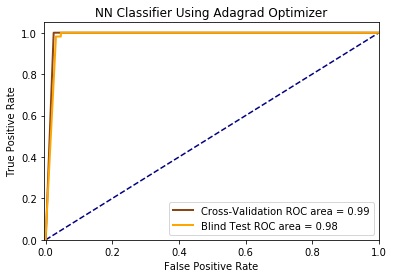}
	\centerline{~~~(e) ~~~~~~~~~~~~~~~~~~~~~~~~~~(f)}
	\includegraphics[width=40mm]{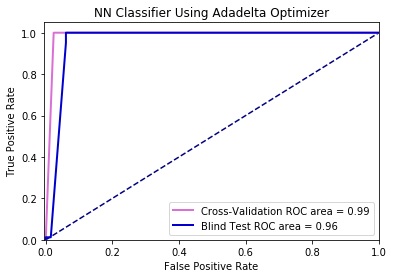}
	\centerline{(g)}
	\caption{ROC plot for 2 class classification of posture; lying down versus rest using (a) Adamax, (b) Adam, (c) sgd, (d) RMSprop, (e) Nadam, (f) Adagrad and (g) Adadelta optimizers.}
	\label{fig_20}
\end{figure}

Further, we used 3 layered ANN with different optimizers to check the accuracy for posture estimation for both 2 class and multi-class classification obtained for both cross-validation and blind test. The confusion matrices obtained from 2 class classification for cross-validation and blind test are shown in the Tables \ref{Tab_IIV}, \ref{Tab_IV}, and \ref{Tab_IVI}. The cross-validation and blind test accuracy of obtained from all the optimizers are shown in Figures \ref{fig_2ClassPosCrossANN} and \ref{fig_2ClassPosBlindANN} respectively. The corresponding ROC plots and their accuracies are shown in Figures \ref{fig_18}, \ref{fig_19}, and \ref{fig_20}.

\begin{table}[h!]
	\renewcommand{\arraystretch}{1.1}
	\setlength{\tabcolsep}{5pt}
	\centering 
	\caption{Cross-Validation and blind test confusion matrix for 2 class  classification of posture, standing versus rest using different classifiers}
	\begin{tabular}{|c|c|c|}
		\hline
		\textbf{Optimizer}&\textbf{Cross-Validation}&\textbf{Blind Test}\\\hline
		Adamax   &	$\begin{bmatrix}82&0\\1&147
		\end{bmatrix}$	&	$\begin{bmatrix}
			79&5\\5&141
		\end{bmatrix}$		\\\hline
		
		Adadelta   &	$\begin{bmatrix}73&0\\0&157
		\end{bmatrix}$	&	$\begin{bmatrix}
			81&3\\3&143
		\end{bmatrix}$		\\\hline
		
		Adam   &	$\begin{bmatrix}73&0\\0&157
		\end{bmatrix}$	&	$\begin{bmatrix}
			79&5\\6&140
		\end{bmatrix}$		\\\hline
		
		Adagrad  &	$\begin{bmatrix}73&0\\0&157
		\end{bmatrix}$	&	$\begin{bmatrix}
			80&4\\4&142
		\end{bmatrix}$		\\\hline
		
		Nadam   &	$\begin{bmatrix}73&0\\0&157
		\end{bmatrix}$	&	$\begin{bmatrix}
			70&3\\2&155
		\end{bmatrix}$		\\\hline
		
		SGD  &	$\begin{bmatrix}73&0\\0&157
		\end{bmatrix}$	&	$\begin{bmatrix}
			72&1\\0&157
		\end{bmatrix}$		\\\hline
		
		RMSprop   &	$\begin{bmatrix}73&0\\0&157
		\end{bmatrix}$	&	$\begin{bmatrix}
			72&1\\1&156
		\end{bmatrix}$		\\\hline
	\end{tabular}
		\label{Tab_IIV}
\end{table}

\begin{table}[h!]
	\renewcommand{\arraystretch}{1.1}
	\setlength{\tabcolsep}{5pt}
	\centering 
	\caption{Cross-Validation and blind test confusion matrix for 2 class  classification of posture, sitting versus rest using different classifiers}
	\begin{tabular}{|c|c|c|}
		\hline
		\textbf{Optimizer}&\textbf{Cross-Validation}&\textbf{Blind Test}\\\hline
		Adamax   &	$\begin{bmatrix}80&1\\0&149
		\end{bmatrix}$	&	$\begin{bmatrix}
			63&7\\7&153
		\end{bmatrix}$		\\\hline
		
		Adadelta   &	$\begin{bmatrix}86&1\\0&143
		\end{bmatrix}$	&	$\begin{bmatrix}
			81&9\\4&136
		\end{bmatrix}$		\\\hline
		
		Adam   &	$\begin{bmatrix}80&1\\1&148
		\end{bmatrix}$	&	$\begin{bmatrix}
			64&6\\7&153
		\end{bmatrix}$		\\\hline
		
		Adagrad  &	$\begin{bmatrix}86&1\\0&143
		\end{bmatrix}$	&	$\begin{bmatrix}
			84&6\\2&138
		\end{bmatrix}$		\\\hline
		
		Nadam   &	$\begin{bmatrix}86&1\\0&143
		\end{bmatrix}$	&	$\begin{bmatrix}
			84&6\\3&137
		\end{bmatrix}$		\\\hline
		
		SGD  &	$\begin{bmatrix}87&0\\0&143
		\end{bmatrix}$	&	$\begin{bmatrix}
			85&5\\10&130
		\end{bmatrix}$		\\\hline
		
		RMSprop   &	$\begin{bmatrix}87&0\\0&143
		\end{bmatrix}$	&	$\begin{bmatrix}
			85&5\\7&133
		\end{bmatrix}$		\\\hline
	\end{tabular}
	\label{Tab_IV}
\end{table}

\begin{table}[h!]
	\renewcommand{\arraystretch}{1.1}
	\setlength{\tabcolsep}{5pt}
	\centering 
	\caption{Cross-Validation and blind test confusion matrix for 2 class  classification of posture, lying down versus rest using different classifiers}
	\begin{tabular}{|c|c|c|}
		\hline
		\textbf{Optimizer}&\textbf{Cross-Validation}&\textbf{Blind Test}\\\hline
		Adamax   &	$\begin{bmatrix}78&0\\0&152
		\end{bmatrix}$	&	$\begin{bmatrix}
			59&6\\6&159
		\end{bmatrix}$		\\\hline
		
		Adadelta   &	$\begin{bmatrix}82&2\\0&146
		\end{bmatrix}$	&	$\begin{bmatrix}
			62&4\\0&164
		\end{bmatrix}$		\\\hline
		
		Adam   &	$\begin{bmatrix}78&0\\0&152
		\end{bmatrix}$	&	$\begin{bmatrix}
			57&8\\7&158
		\end{bmatrix}$		\\\hline
		
		Adagrad  &	$\begin{bmatrix}82&2\\0&146
		\end{bmatrix}$	&	$\begin{bmatrix}
			62&4\\0&164
		\end{bmatrix}$		\\\hline
		
		Nadam   &	$\begin{bmatrix}83&1\\2&144
		\end{bmatrix}$	&	$\begin{bmatrix}
			63&3\\0&164
		\end{bmatrix}$		\\\hline
		
		SGD  &	$\begin{bmatrix}84&0\\0&146
		\end{bmatrix}$	&	$\begin{bmatrix}
			61&5\\0&164
		\end{bmatrix}$		\\\hline
		
		RMSprop   &	$\begin{bmatrix}84&0\\0&146
		\end{bmatrix}$	&	$\begin{bmatrix}
			61&5\\3&161
		\end{bmatrix}$		\\\hline
	\end{tabular}
	\label{Tab_IVI}
\end{table}


\begin{table}
	\renewcommand{\arraystretch}{1.1}
	\setlength{\tabcolsep}{2pt}
	\caption{Accuracy and confusion matrix for multi-class classification  of posture using ANN with different optimizer}
	\centering 
	\label{Tab_IVII}
	\begin{tabular}{|c|c|c|c|c|}\hline
		\textbf{Optimizer}&\multicolumn{2}{c|}{\textbf{Cross-Validation}}&\multicolumn{2}{c|}{\textbf{Blind Test}}\\\hline
		&\textbf{Accuracy}&\textbf{Confusion}&\textbf{Accuracy}&\textbf{Confusion}\\
		&\textbf{\%}&\textbf{Matrix}&\textbf{\%}&\textbf{Matrix}\\\hline
		
		Adamax&100&$\begin{bmatrix}
			73&0&0\\0&81&0\\0&0&77
		\end{bmatrix}$&96.62&$\begin{bmatrix}
		73&0&0\\7&74&0\\2&2&73
	\end{bmatrix}$\\\hline

Adadelta&100&$\begin{bmatrix}
	73&0&0\\0&81&0\\0&0&77
\end{bmatrix}$&94.37&$\begin{bmatrix}
72&1&0\\10&71&0\\1&5&71
\end{bmatrix}$\\\hline

Adam&100&$\begin{bmatrix}
		73&0&0\\0&81&0\\0&1&77
\end{bmatrix}$&91.73&$\begin{bmatrix}
72&0&1\\9&71&1\\0&6&71
\end{bmatrix}$\\\hline

Adagrad&99.35&$\begin{bmatrix}
	73&0&0\\0&81&0\\0&2&75
\end{bmatrix}$&96.28&$\begin{bmatrix}
70&0&3\\5&76&0\\2&2&73
\end{bmatrix}$\\\hline

Nadam&100&$\begin{bmatrix}
	73&0&0\\0&81&0\\0&0&77
\end{bmatrix}$&96.54&$\begin{bmatrix}
73&0&0\\4&74&3\\3&2&72
\end{bmatrix}$\\\hline

SGD&100&$\begin{bmatrix}
	73&0&0\\1&80&0\\0&0&77
\end{bmatrix}$&92.77&$\begin{bmatrix}
70&1&2\\6&71&4\\3&6&68
\end{bmatrix}$\\\hline

RMSprop&100&$\begin{bmatrix}
	73&0&0\\0&81&0\\0&1&77
\end{bmatrix}$&93.81&$\begin{bmatrix}
71&0&2\\6&72&3\\4&3&70
\end{bmatrix}$\\\hline
		
	\end{tabular}
\end{table}

The multi-class classification of posture using different optimizers obtained from 3 layered ANN. The results obtained from various optimize are checked for multi-class classification in terms of accuracy and confusion matrix which are listed in Table \ref{Tab_IVII}. Here we see that maximum cross-validation accuracy of 100\% is obtained from almost all the optimizers except Adagrad. Whereas, maximum blind test accuracy of 96.62\% is obtained from Adamax optimizer for multi-class classification of posture.

\begin{figure}[h!]
	\centering
	\includegraphics[width=50mm]{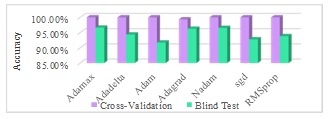}
	\caption{Accuracy for multiclass classification of posture using ANN with different optimizers.}
	\label{fig_21}
\end{figure}

On using 2-class classification, to know the exact posture of the occupant, it is required to check for three times (i.e. for ‘standing versus rest’, ‘sitting versus rest’ and ‘lying down versus rest’). This is more time consuming and so we finally used multiclass classification for posture determination. The maximum blind test accuracy obtained from the random forest classifier is 98.27\% and is used for illumination regulation.

\subsection{Emotion recognition}

\begin{figure}[h!]
	\centering
	\includegraphics[width=40mm]{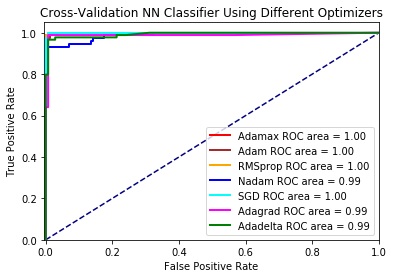}
	\includegraphics[width=40mm]{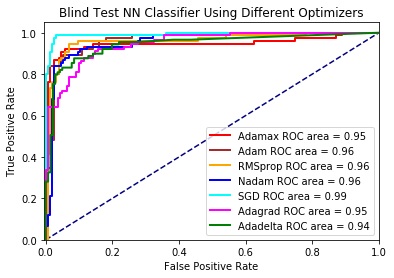}
	\centerline{~~~(a) ~~~~~~~~~~~~~~~~~~~~~~~~~~(b)}
	\includegraphics[width=40mm]{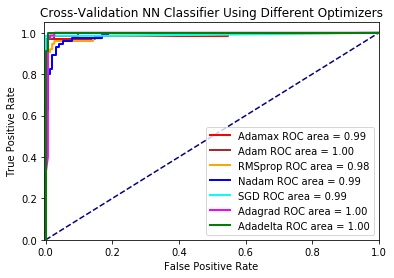}
	\includegraphics[width=40mm]{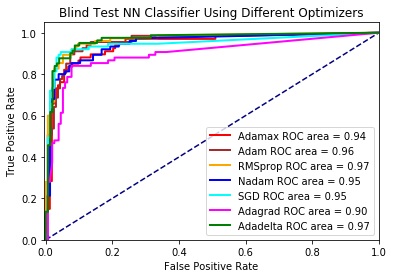}
	\centerline{~~~(c) ~~~~~~~~~~~~~~~~~~~~~~~~~~(d)}
	\includegraphics[width=40mm]{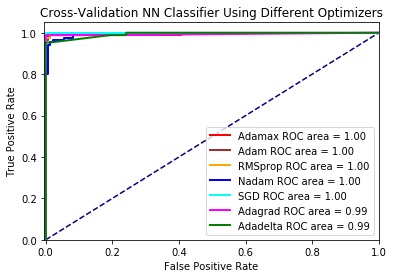}
	\includegraphics[width=40mm]{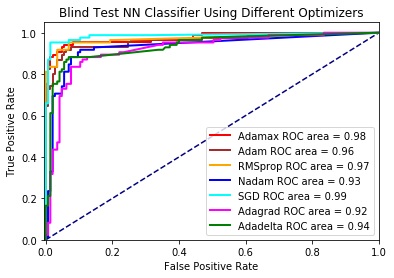}
	\centerline{~~~(e) ~~~~~~~~~~~~~~~~~~~~~~~~~~(f)}
	\caption{ROC of 2 class classification; for comfortable vs. rest (a) cross-validation (b) blind test; for neutral vs. rest (c) cross-validation (d) blind test and for uncomfortable vs. rest (e) cross-validation (f) blind test to determine emotion.}
	\label{fig_22}
\end{figure}

Face point data are obtained for $31$ people using Visual Studio 2019 (WPF application C\#). A neural network training using deep learning technique is done using different optimizers for predicting emotion which is classified as `comfortable', `neutral' and `uncomfortable'. The input layer of the neural network has $46$ nodes, after which there are $4$ hidden layers with decreasing number of nodes i.e. from $50$ to $20$ by $10$. Then the output from the $4$-th hidden layer is dropped by $20\%$ which is again passed through a $5$-th hidden layer which has $20$ nodes. Output from the $5$-th hidden layer is again dropped by 30\%. Thereafter passing that through $5$ hidden layers with number of nodes decreasing and then again slightly increasing i.e. $20$ to $10$ to $5$ to $10$ to $20$. Output from the 6th hidden layer and the $10$-th hidden layer is merged. And the merged set is again passed through $2$ hidden layers with nodes $20$ and $25$. The output from the 4th hidden layer and the $12$-th hidden layer is again merged. The merged set is passed through $2$ hidden layers with nodes $25$ and $7$. Finally the output from this $14$-th hidden layer is fed to the output layer which has a single node. Therefore, overall, we have $14$ hidden layers. All the hidden layers as rectified linear unit activation function.
Both 2 class and multi-class classification is done. In 2 class classification of emotion area under the ROC plots varies from $0.9$ to $1$ for different optimizers as shown in Figure \ref{fig_22} and their corresponding confusion matrices are provided in Tables \ref{Tab_IVIII}, \ref{Tab_IIX}, and \ref{Tab_IX}.

\begin{table}[h!]
	\renewcommand{\arraystretch}{1.1}
	\setlength{\tabcolsep}{5pt}
	\centering 
	\caption{Confusion matrix for 2 class classification of comfortable versus rest}
	\begin{tabular}{|c|c|c|}
		\hline
		\textbf{Optimizer}&\textbf{Cross-Validation}&\textbf{Blind Test}\\\hline
		Adamax   &	$\begin{bmatrix}155&0\\1&74
		\end{bmatrix}$	&	$\begin{bmatrix}
			148&7\\7&68
		\end{bmatrix}$		\\\hline
		
		Adadelta   &	$\begin{bmatrix}140&1\\3&86
		\end{bmatrix}$	&	$\begin{bmatrix}
			134&7\\15&74
		\end{bmatrix}$		\\\hline
		
		Adam   &	$\begin{bmatrix}154&1\\2&73
		\end{bmatrix}$	&	$\begin{bmatrix}
			140&15\\8&67
		\end{bmatrix}$		\\\hline
		
		Adagrad  &	$\begin{bmatrix}140&1\\1&88
		\end{bmatrix}$	&	$\begin{bmatrix}
			132&9\\23&66
		\end{bmatrix}$		\\\hline
		
		Nadam   &	$\begin{bmatrix}155&0\\5&70
		\end{bmatrix}$	&	$\begin{bmatrix}
			146&9\\10&65
		\end{bmatrix}$		\\\hline
		
		SGD  &	$\begin{bmatrix}155&0\\2&73
		\end{bmatrix}$	&	$\begin{bmatrix}
			152&3\\4&71
		\end{bmatrix}$		\\\hline
		
		RMSprop   &	$\begin{bmatrix}154&1\\1&74
		\end{bmatrix}$	&	$\begin{bmatrix}
			146&9\\8&67
		\end{bmatrix}$		\\\hline
	\end{tabular}
	\label{Tab_IVIII}
\end{table}

\begin{table}[h!]
	\renewcommand{\arraystretch}{1.1}
	\setlength{\tabcolsep}{5pt}
	\centering 
	\caption{Confusion matrix for 2 class classification of neutral versus rest}
	\begin{tabular}{|c|c|c|}
		\hline
		\textbf{Optimizer}&\textbf{Cross-Validation}&\textbf{Blind Test}\\\hline
		Adamax   &	$\begin{bmatrix}163&0\\2&65
		\end{bmatrix}$	&	$\begin{bmatrix}
			153&10\\12&55
		\end{bmatrix}$		\\\hline
		
		Adadelta   &	$\begin{bmatrix}148&1\\0&81
		\end{bmatrix}$	&	$\begin{bmatrix}
			143&6\\11&70
		\end{bmatrix}$		\\\hline
		
		Adam   &	$\begin{bmatrix}160&3\\2&65
		\end{bmatrix}$	&	$\begin{bmatrix}
			153&10\\11&56
		\end{bmatrix}$		\\\hline
		
		Adagrad  &	$\begin{bmatrix}154&1\\2&73
		\end{bmatrix}$	&	$\begin{bmatrix}
			145&10\\14&61
		\end{bmatrix}$		\\\hline
		
		Nadam   &	$\begin{bmatrix}150&5\\5&70
		\end{bmatrix}$	&	$\begin{bmatrix}
			146&9\\10&65
		\end{bmatrix}$		\\\hline
		
		SGD  &	$\begin{bmatrix}163&0\\1&66
		\end{bmatrix}$	&	$\begin{bmatrix}
			149&9\\9&66
		\end{bmatrix}$		\\\hline
		
		RMSprop   &	$\begin{bmatrix}150&5\\3&72
		\end{bmatrix}$	&	$\begin{bmatrix}
			143&12\\6&69
		\end{bmatrix}$		\\\hline
	\end{tabular}
	\label{Tab_IIX}
\end{table}

\begin{table}[h!]
	\renewcommand{\arraystretch}{1.1}
	\setlength{\tabcolsep}{5pt}
	\centering 
	\caption{Confusion matrix for 2 class classification of uncomfortable versus rest}
	\begin{tabular}{|c|c|c|}
		\hline
		\textbf{Optimizer}&\textbf{Cross-Validation}&\textbf{Blind Test}\\\hline
		Adamax   &	$\begin{bmatrix}145&0\\1&84
		\end{bmatrix}$	&	$\begin{bmatrix}
			141&4\\0&76
		\end{bmatrix}$		\\\hline
		
		Adadelta   &	$\begin{bmatrix}145&0\\4&81
		\end{bmatrix}$	&	$\begin{bmatrix}
			139&6\\16&69
		\end{bmatrix}$		\\\hline
		
		Adam   &	$\begin{bmatrix}144&1\\1&84
		\end{bmatrix}$	&	$\begin{bmatrix}
			138&7\\10&75
		\end{bmatrix}$		\\\hline
		
		Adagrad  &	$\begin{bmatrix}145&0\\1&84
		\end{bmatrix}$	&	$\begin{bmatrix}
			133&12\\14&71
		\end{bmatrix}$		\\\hline
		
		Nadam   &	$\begin{bmatrix}143&2\\4&81
		\end{bmatrix}$	&	$\begin{bmatrix}
			134&11\\11&74
		\end{bmatrix}$		\\\hline
		
		SGD  &	$\begin{bmatrix}141&4\\0&85
		\end{bmatrix}$	&	$\begin{bmatrix}
			143&2\\6&79
		\end{bmatrix}$		\\\hline
		
		RMSprop   &	$\begin{bmatrix}144&1\\2&83
		\end{bmatrix}$	&	$\begin{bmatrix}
			140&5\\9&76
		\end{bmatrix}$		\\\hline
	\end{tabular}
	\label{Tab_IX}
\end{table}

\begin{figure}[h!]
	\centering
	\includegraphics[width=50mm]{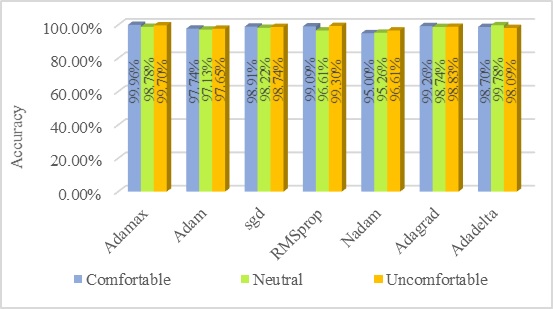}
	\caption{Cross-validation accuracy plot for 2 class classification of emotion.}
	\label{fig_23}
\end{figure}
\begin{figure}[h!]
	\centering
	\includegraphics[width=50mm]{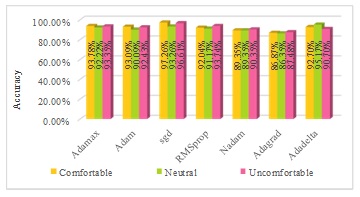}
	\caption{Blind test accuracy plot for 2 class classification of emotion.}
	\label{fig_24}
\end{figure}

\begin{table}
		\renewcommand{\arraystretch}{1.1}
	\setlength{\tabcolsep}{2pt}
	\centering 
	\caption{Confusion matrix of multi-class classification to determine emotion}
	\begin{tabular}{|c|c|c|c|c|}\hline
		\textbf{Optimizer}&\multicolumn{2}{c|}{\textbf{Cross-Validation}}&\multicolumn{2}{c|}{\textbf{Blind Test}}\\\cline{2-5}
		&\textbf{Accuracy}&\textbf{Confusion}&\textbf{Accuracy}&\textbf{Confusion}\\
		&\textbf{\%}&\textbf{Matrix}&\textbf{\%}&\textbf{Matrix}\\\hline
		
		Adamax&99.13&$\begin{bmatrix}
			85&1&0\\1&70&0\\0&0&73
		\end{bmatrix}$&92.57&$\begin{bmatrix}
		76&7&3\\7&59&5\\0&2&71
	\end{bmatrix}$\\\hline

Adadelta&97.70&$\begin{bmatrix}
	82&2&2\\1&69&1\\0&0&73
\end{bmatrix}$&94.13&$\begin{bmatrix}
80&3&2\\5&62&4\\0&1&72
\end{bmatrix}$\\\hline

		Adam&97.70&$\begin{bmatrix}
			83&1&2\\2&68&1\\0&1&72
		\end{bmatrix}$&90&$\begin{bmatrix}
		74&8&4\\4&55&12\\3&2&68
	\end{bmatrix}$\\\hline

Adagrad&98.91&$\begin{bmatrix}
	83&3&0\\0&70&1\\0&1&72
\end{bmatrix}$&74.61&$\begin{bmatrix}
70&14&2\\8&55&8\\5&2&66
\end{bmatrix}$\\\hline

Nadam&90.96&$\begin{bmatrix}
	75&5&6\\6&56&9\\2&2&69
\end{bmatrix}$&82.61&$\begin{bmatrix}
73&12&1\\6&50&5\\1&14&58
\end{bmatrix}$\\\hline

SGD&95.13&$\begin{bmatrix}
	84&2&0\\1&65&5\\0&2&71
\end{bmatrix}$&91.26&$\begin{bmatrix}
81&5&0\\2&67&2\\0&12&61
\end{bmatrix}$\\\hline

RMSprop&97.74&$\begin{bmatrix}
	84&2&0\\0&71&0\\0&3&70
\end{bmatrix}$&88.39&$\begin{bmatrix}
82&4&0\\8&57&6\\1&7&65
\end{bmatrix}$\\\hline
	\end{tabular}
	\label{Tab_IXI}
\end{table}

\begin{figure}[h!]
	\centering
	\includegraphics[width=50mm]{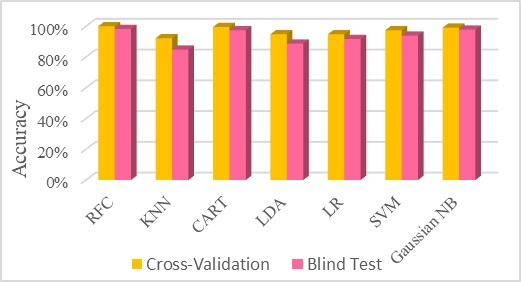}
	\caption{Multiclass classification accuracy for posture determination.}
	\label{fig_MultiClassPos}
\end{figure}

In case the of multi-class classification maximum cross-validation accuracy is obtained for Adamax and maximum blind test accuracy is obtained for Adadelta optimizer as shown in Table \ref{Tab_IXI} and Figure \ref{fig_MultiClassPos}. Thus for emotion detection finally we use Adadelta which gives the maximum blind test accuracy of $94.13\%$ for multi-class classification. Similarly, as we mentioned in case of posture detection we used multi-class classification as the 2-class classification needs to be checked three times (i.e. `comfortable versus rest', `neutral versus rest' and `uncomfortable versus rest') to detect the exact emotion requiring more time for computation.

\subsection{Final code of automation}
\begin{figure}[h!]
	\centering
	\includegraphics[width=70mm]{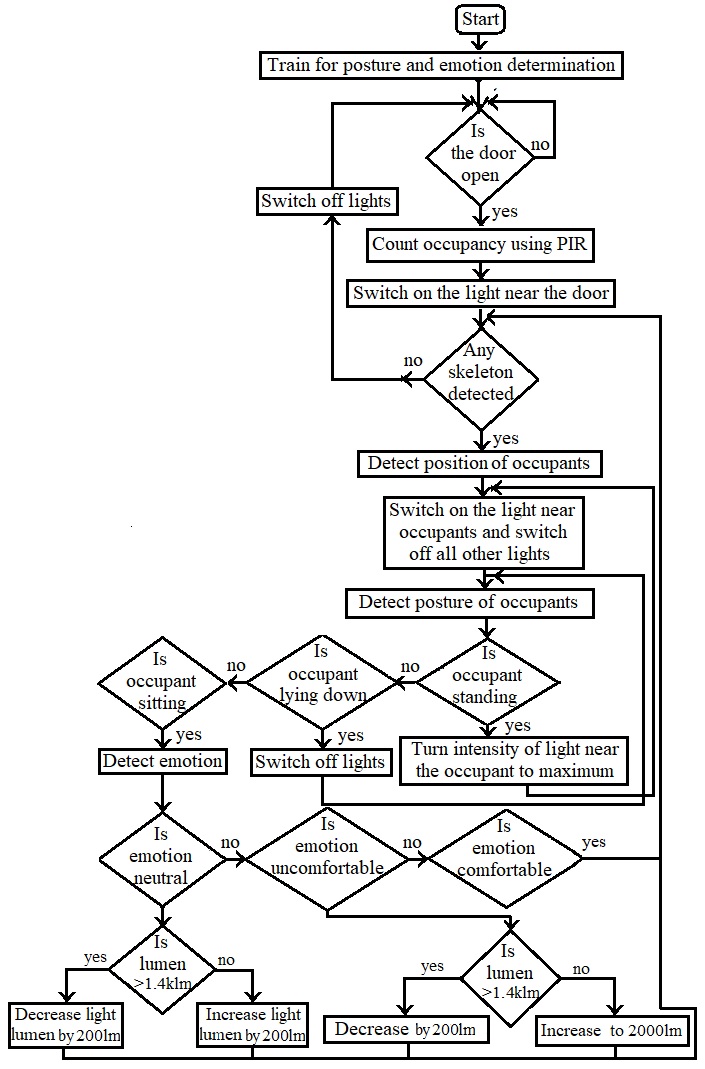}
	\caption{Flowchart for automated change in illumination.}
	\label{fig_Flowchart}
\end{figure}

While the system runs in automation following the flowchart shown in Figure \ref{fig_Flowchart}, multi-class classification is performed using RFC to predict posture and Adadelta optimizer to predict emotion. Data from a door sensor, PIR sensors and BH1750 are sent wirelessly over socket communication using Raspberry Pi Zero W.
\section{Conclusion}
We proposed a scheme for regulating the illumination of a room based on the posture and comfort level of the occupant. A sensor network is built that provides information on the presence of an occupant, and the occupant’s coordinate position of joints and facial points to determine posture and emotion using machine learning and correspondingly change the illumination of the room, and the illumination is also sensed using illumination sensor. The sensor network works wirelessly using Raspberry Pi Zero W. Thus preventing unnecessary energy wastage due to needless illumination in the room. 
This requires accurate detection of posture and emotion due to which we created an efficient framework and performed several machine learning techniques. Out of which, maximum accuracy of $98.27\%$ was obtained on a blind test from the random forest classifier for the multi-class classification. Deep learning is used to detect emotion where various optimizers were used out of which Adadelta performed with maximum accuracy of $94.13\%$ on a blind test for the multi-class classification.
\section{Future Work}
The system we prepared was for a single occupant room as the Kinect XBOX 360 we used can properly detect coordinates of only a single occupant. On using XBOX 1 the same scheme of illumination regulation can be performed for multi-occupant which is planned as future work. The occupancy modeling we performed here is done only by changing the illumination in the room. Other electrical equipment must be regulated like air-conditioners, fans, \emph{etc.,} \cite{yang2014systematic} to maintain occupant’s comfort and reduce energy wastage. The use of a control scheme may provide the desired illumination more efficiently \cite{copot2018pid}.  
%	%%=============================================================	
\bibliographystyle{ieeetr}
\bibliography{Bibliography}

\begin{thebibliography}{10}

\bibitem{chen2015modeling}
Z.~Chen, J.~Xu, and Y.~C. Soh, ``Modeling regular occupancy in commercial
  buildings using stochastic models,'' {\em Energy and Buildings}, vol.~103,
  pp.~216--223, 2015.

\bibitem{liao2010integrated}
C.~Liao and P.~Barooah, ``An integrated approach to occupancy modeling and
  estimation in commercial buildings,'' in {\em Proceedings of the 2010
  American control conference}, pp.~3130--3135, IEEE, 2010.

\bibitem{wren2007merl}
C.~R. Wren, Y.~A. Ivanov, D.~Leigh, and J.~Westhues, ``The merl motion detector
  dataset,'' in {\em Proceedings of the 2007 workshop on Massive datasets},
  pp.~10--14, 2007.

\bibitem{erickson2009energy}
V.~L. Erickson, Y.~Lin, A.~Kamthe, R.~Brahme, A.~Surana, A.~E. Cerpa, M.~D.
  Sohn, and S.~Narayanan, ``Energy efficient building environment control
  strategies using real-time occupancy measurements,'' in {\em Proceedings of
  the first ACM workshop on embedded sensing systems for energy-efficiency in
  buildings}, pp.~19--24, 2009.

\bibitem{moon2010ann}
J.~W. Moon and J.-J. Kim, ``Ann-based thermal control models for residential
  buildings,'' {\em Building and Environment}, vol.~45, no.~7, pp.~1612--1625,
  2010.

\bibitem{arief2017cd}
I.~B. Arief-Ang, F.~D. Salim, and M.~Hamilton, ``Cd-hoc: indoor human occupancy
  counting using carbon dioxide sensor data,'' {\em arXiv preprint
  arXiv:1706.05286}, 2017.

\bibitem{maaspuro2018infrared}
M.~Maaspuro, ``Infrared occupancy detection technologies in building
  automation-a review,'' {\em Journal of Engineering and Applied Sciences
  (ARPN)}, vol.~13, no.~19, pp.~8055--8068, 2018.

\bibitem{bahl2000radar}
P.~Bahl and V.~N. Padmanabhan, ``Radar: An in-building rf-based user location
  and tracking system,'' in {\em Proceedings IEEE INFOCOM 2000. Conference on
  computer communications. Nineteenth annual joint conference of the IEEE
  computer and communications societies (Cat. No. 00CH37064)}, vol.~2,
  pp.~775--784, Ieee, 2000.

\bibitem{gu2009survey}
Y.~Gu, A.~Lo, and I.~Niemegeers, ``A survey of indoor positioning systems for
  wireless personal networks,'' {\em IEEE Communications surveys \& tutorials},
  vol.~11, no.~1, pp.~13--32, 2009.

\bibitem{liu2007survey}
H.~Liu, H.~Darabi, P.~Banerjee, and J.~Liu, ``Survey of wireless indoor
  positioning techniques and systems,'' {\em IEEE Transactions on Systems, Man,
  and Cybernetics, Part C (Applications and Reviews)}, vol.~37, no.~6,
  pp.~1067--1080, 2007.

\bibitem{domdouzis2007radio}
K.~Domdouzis, B.~Kumar, and C.~Anumba, ``Radio-frequency identification (rfid)
  applications: A brief introduction,'' {\em Advanced Engineering Informatics},
  vol.~21, no.~4, pp.~350--355, 2007.

\bibitem{vujovic2014raspberry}
V.~Vujovi{\'c} and M.~Maksimovi{\'c}, ``Raspberry pi as a wireless sensor node:
  Performances and constraints,'' in {\em 2014 37th international convention on
  information and communication technology, electronics and microelectronics
  (MIPRO)}, pp.~1013--1018, IEEE, 2014.

\bibitem{ferdoush2014wireless}
S.~Ferdoush and X.~Li, ``Wireless sensor network system design using raspberry
  pi and arduino for environmental monitoring applications,'' {\em Procedia
  Computer Science}, vol.~34, pp.~103--110, 2014.

\bibitem{yang2014systematic}
Z.~Yang, N.~Li, B.~Becerik-Gerber, and M.~Orosz, ``A systematic approach to
  occupancy modeling in ambient sensor-rich buildings,'' {\em Simulation},
  vol.~90, no.~8, pp.~960--977, 2014.

\bibitem{munir2017real}
S.~Munir, R.~S. Arora, C.~Hesling, J.~Li, J.~Francis, C.~Shelton, C.~Martin,
  A.~Rowe, and M.~Berges, ``Real-time fine grained occupancy estimation using
  depth sensors on arm embedded platforms,'' in {\em 2017 IEEE Real-Time and
  Embedded Technology and Applications Symposium (RTAS)}, pp.~295--306, IEEE,
  2017.

\bibitem{gordon2002wireless}
D.~D. Gordon-Levitt and D.~Carner, ``Wireless reed switch-based burglar
  alarm,'' June~4 2002.
\newblock US Patent 6,400,267.

\bibitem{canceill1969magnetic}
B.~J.-J. Canceill, ``Magnetic closure and switch for doors and similar
  devices,'' Feb.~4 1969.
\newblock US Patent 3,426,166.

\bibitem{kaundanya2017smart}
C.~Kaundanya, O.~Pathak, A.~Nalawade, and S.~Parode, ``Smart surveillance
  system using raspberry pi and face recognition,'' {\em International Journal
  of Advanced Research in Computer and Communication Engineering}, vol.~6,
  no.~4, pp.~622--623, 2017.

\bibitem{patel2016smart}
P.~B. Patel, V.~M. Choksi, S.~Jadhav, and M.~Potdar, ``Smart motion detection
  system using raspberry pi,'' {\em International Journal of Applied
  Information Systems (IJAIS)}, vol.~10, no.~5, pp.~37--40, 2016.

\bibitem{gao2017light}
J.~Gao, J.~Luo, A.~Xu, and J.~Yu, ``Light intensity intelligent control system
  research and design based on automobile sun visor of bh1750,'' in {\em 2017
  29th Chinese Control And Decision Conference (CCDC)}, pp.~3957--3960, IEEE,
  2017.

\bibitem{wang2011design}
J.~Wang, T.-F. Mao, and Y.~Chen, ``Design and implementation of luminometer
  system based on novel bh1750 ic,'' {\em Journal of Changshu Institute of
  Technology}, pp.~586--589, 2011.

\bibitem{copot2018pid}
C.~Copot, T.~Mac~Thi, and C.~Ionescu, ``Pid based particle swarm optimization
  in offices light control,'' {\em IFAC-PapersOnLine}, vol.~51, no.~4,
  pp.~382--387, 2018.

\end{thebibliography}
%	%%==========================================================  
	\end{document}